\documentclass{acm_proc_article-sp}
\usepackage{subcaption}
\usepackage{color}
\definecolor{gray97}{gray}{.97}
\definecolor{gray75}{gray}{.75}
\definecolor{gray45}{gray}{.45}
\usepackage{listings}
\usepackage{comment}

\usepackage{dblfloatfix}

\lstdefinestyle{C}
   {language=C,
   }

\begin{document}
%
\conferenceinfo{HIP3ES}{2015 Amsterdam, The Netherlands}

\title{Reducing overheads of dynamic scheduling on heterogeneous chips 
}
%
%
%
%
%

\numberofauthors{2} 
%
\author{
%
%
\alignauthor
Francisco Corbera, Andr\'es Rodr\'{\i}guez, Rafael Asenjo, Angeles Navarro, Antonio Vilches, 
\\
       \affaddr{Universidad de M\'alaga, Andaluc\'{\i}a Tech, Dept. of
         Computer Architecture, Spain}\\
       \email{\small\{corbera,andres,asenjo,angeles,avilches\}@ac.uma.es}
\and  
\alignauthor
Mar\'{\i}a J. Garzar\'an\\
       \affaddr{Dept. Computer Science, UIUC}\\
       \email{\small garzaran@illinois.edu}
}

\maketitle
\begin{abstract}
In recent processor development, we have witnessed the integration of
GPU and CPUs into a single chip. The result of this integration is a
reduction of the data communication overheads. This enables an
efficient collaboration of both devices in the execution of parallel
workloads.

In this work, we focus on the problem of efficiently scheduling chunks
of iterations of parallel loops among the computing devices on the
chip (the GPU and the CPU cores) in the context of irregular
applications. In particular, we analyze the sources of overhead that
the host thread experiments when a chunk of iterations is offloaded to
the GPU while other threads are executing concurrently other chunks on
the CPU cores. We carefully study these overheads on different
processor architectures and operating systems using Barnes Hut as a
study case representative of irregular applications.  We also propose
a set of optimizations to mitigate the overheads that arise in
presence of oversubscription and take advantage of the different
features of the heterogeneous architectures. Thanks to these
optimizations we reduce Energy-Delay Product (EDP) by 18\% and 84\% on
Intel Ivy Bridge and Haswell architectures, respectively, and by 57\%
on the Exynos big.LITTLE.

\end{abstract}




\section{Introduction}

Recently, we have seen a trend towards the integration of GPU and CPUs
on the same die. Examples include recent Intel processors (Ivy Bridge,
Haswell), the AMD APUs, or more power constrained processors targeted
at mobile and embedded devices, like the Samsung Exynos 5 or the
Qualcomm Snapdragon 800, among others. The integration allows the
sharing of the memory system, what reduces the communication overheads
between the devices and enables a more effective cooperation among all
the computing devices. In contrat to discrete GPUs (connected
through a slower PCI bus), integrated GPUs also enable the
implementation of more dynamic strategies to distribute the workload
among the cores and the GPU.  In fact, achieving maximum performance
and/or minimum energy consumption often requires simultaneous use of
both, the GPU and the CPU cores~\cite{concord}.


In this context, one problem that has received attention lately is the
efficient execution of the iterations of a parallel loop on both
devices, the CPU cores and the integrated GPU.  However, the optimal
division of work between CPU and GPU is very application and input
data dependent so it is
required a careful partitioning of the workload across the CPU cores
and the GPU accelerator. In the case of regular applications, the main
difficulty is to determine how to partition the load between the two
devices to avoid load imbalance. This is usually accomplished by
running a few chunks of iterations in both devices to determine the
speed difference between them.  The appropriate chunk for each device
is then computed and scheduled to run~\cite{Bel13}.  In the case of
irregular applications the challenge is bigger because, in this case,
GPU performance can be suboptimal when the application workload is
distributed among all the cores and the GPU without considering the
size of the chunk assigned to the GPU, even when the load is balanced among the devices.
For instance, using the irregular Barnes Hut benchmark~\cite{BarnesGPU} as our case
of study, we have found that the size of the chunk of iterations
assigned to the GPU has a significant impact on its performance, as we
will show in Section~\ref{sec:motivation}.  Thus, for each
application and input data, the scheduling strategy must be aware of the optimal
chunk of iterations to offload to the GPU.  Since the workload
assigned to the GPU and the CPU cores must also be balanced, the best
approach is a dynamic scheduling strategy that assigns the optimal
chunk of iterations to each device. This means that the cores and the
GPU will repeatedly receive a chunk of iterations to be computed until
the end of the iteration space. This scheduling strategy is described
in Section~\ref{sec:scheduling}.
\begin{figure*}[htb]
        \centering
        \begin{subfigure}[b]{0.3\textwidth}
           \includegraphics[width=\textwidth]{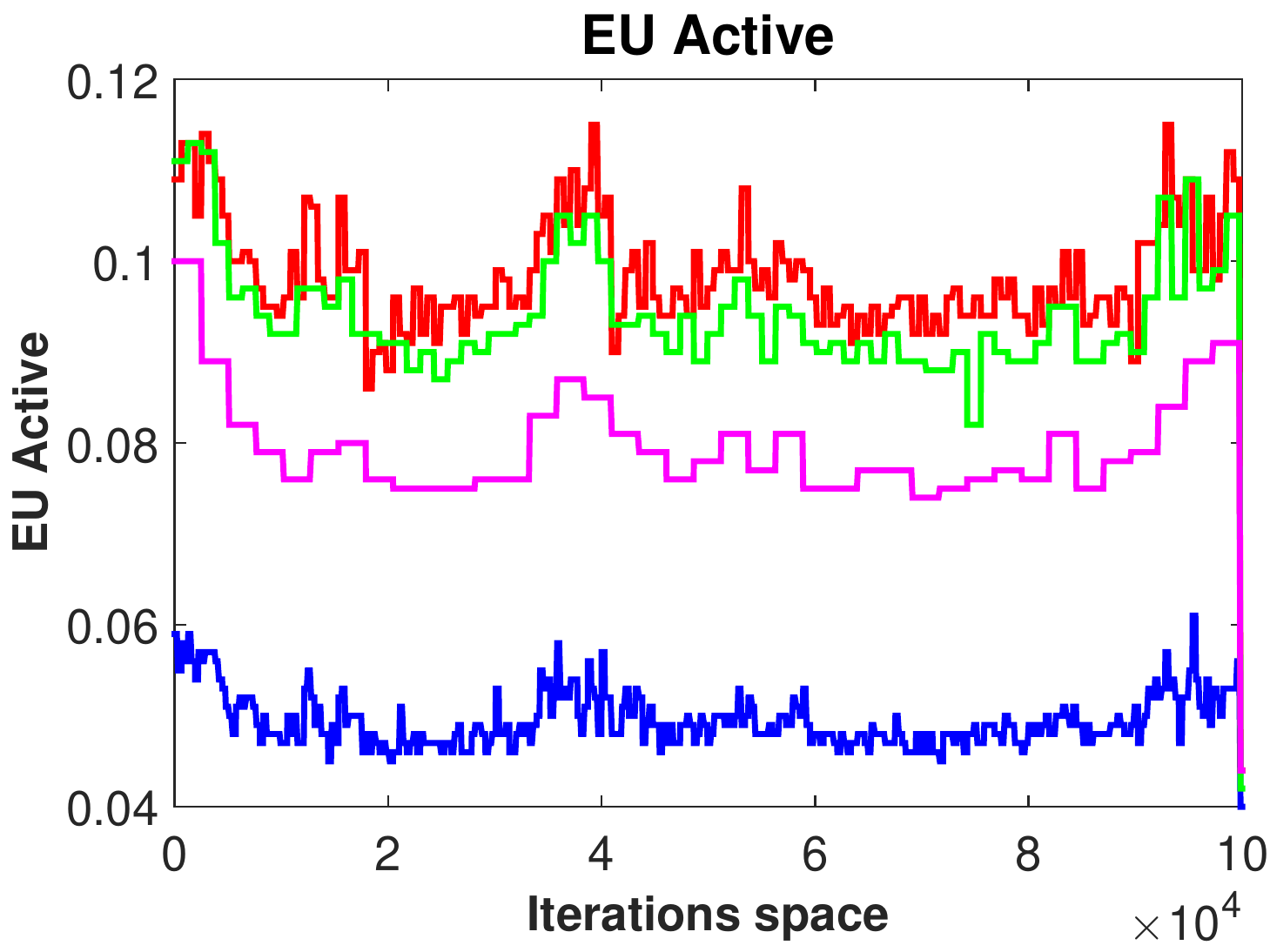}%
           \caption{}\label{fig:bhactive}
        \end{subfigure}%
        ~~ 
        \begin{subfigure}[b]{0.3\textwidth}
           \includegraphics[width=\textwidth]{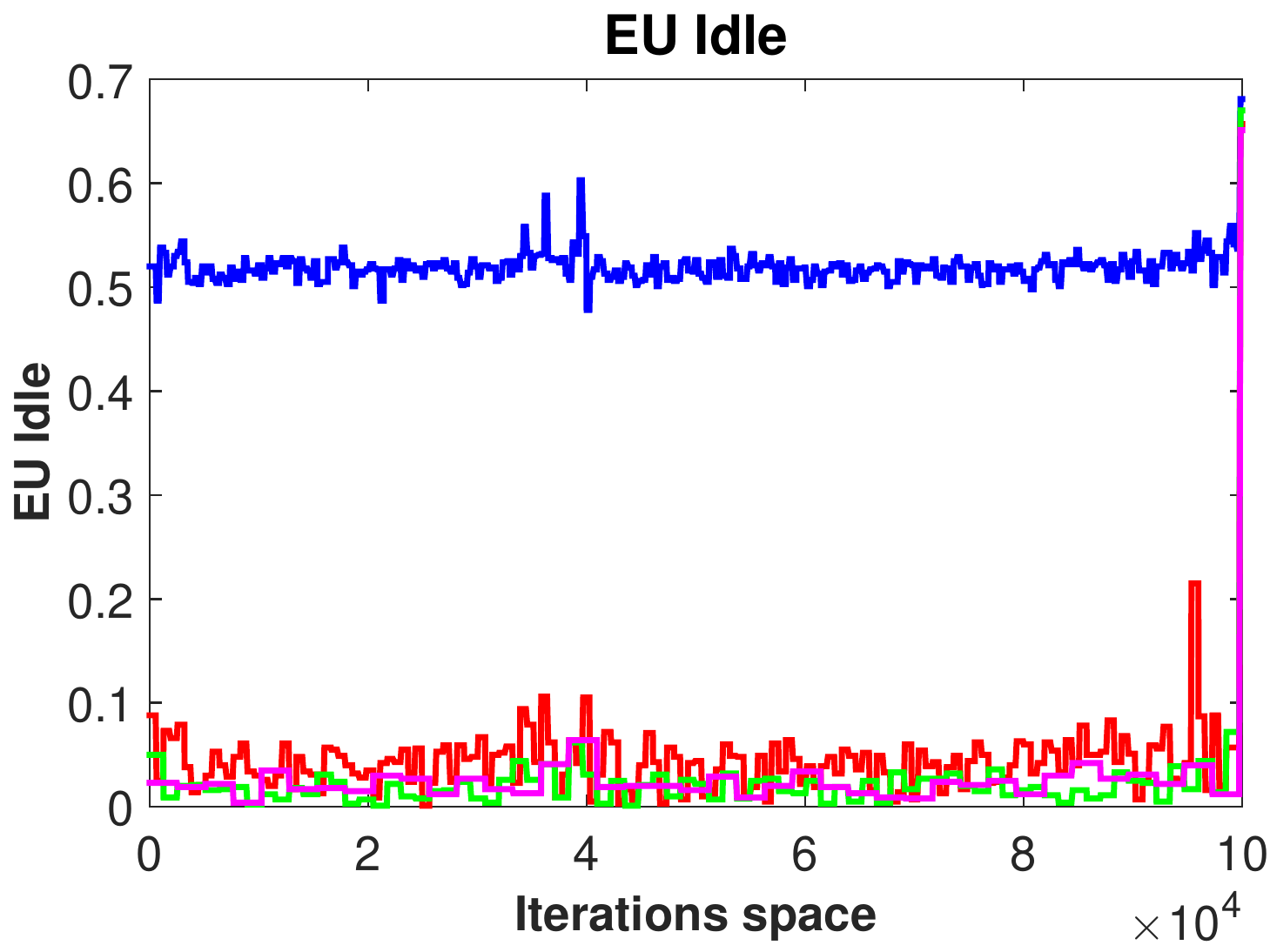}%
           \caption{}\label{fig:bhidle}
        \end{subfigure}
        ~~ 
        \begin{subfigure}[b]{0.3\textwidth}
           \includegraphics[width=\textwidth]{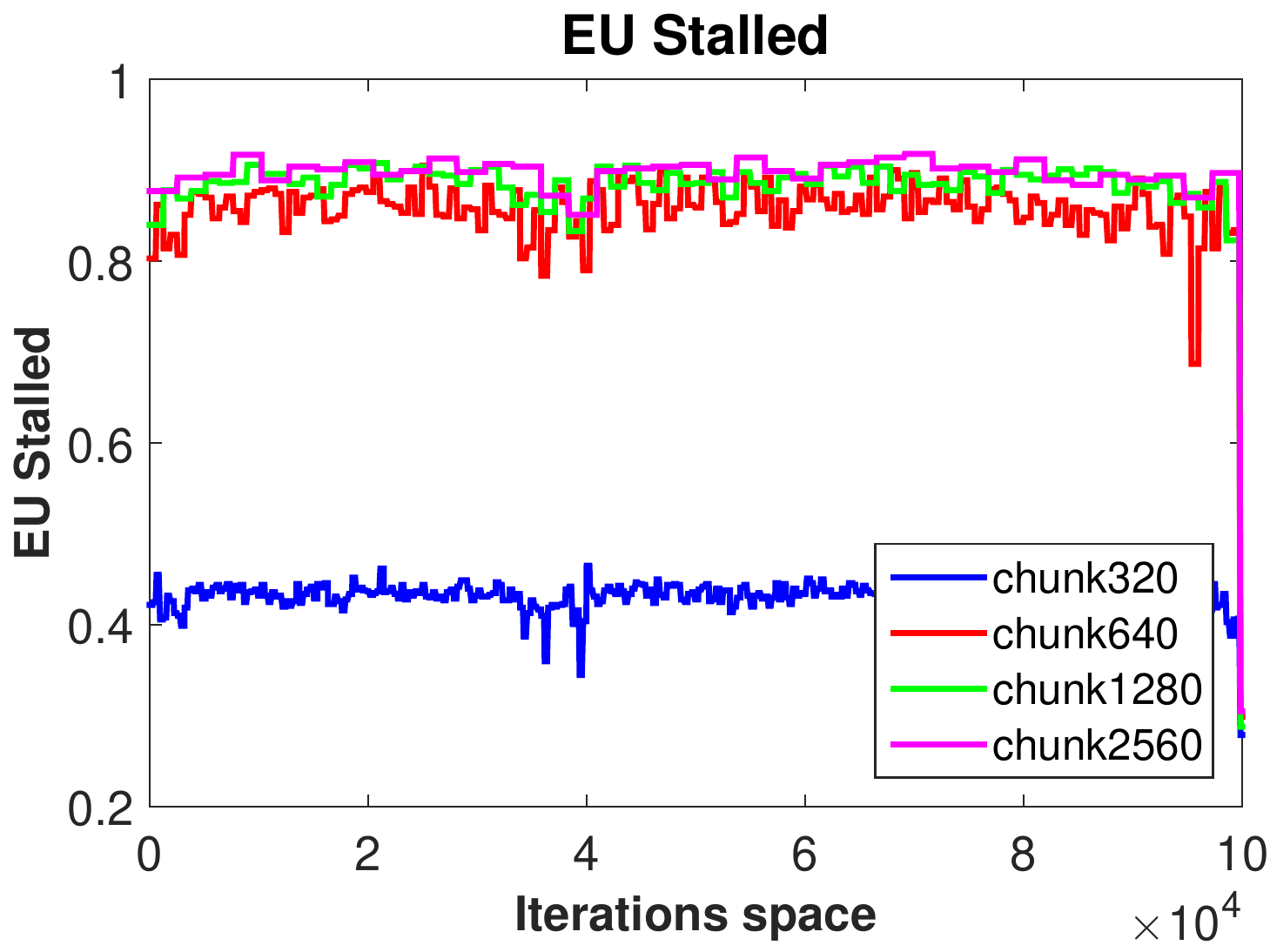}%
           \caption{}\label{fig:bhstalled}
         \end{subfigure}

           \begin{subfigure}[b]{0.33\textwidth}
           \includegraphics[width=\textwidth]{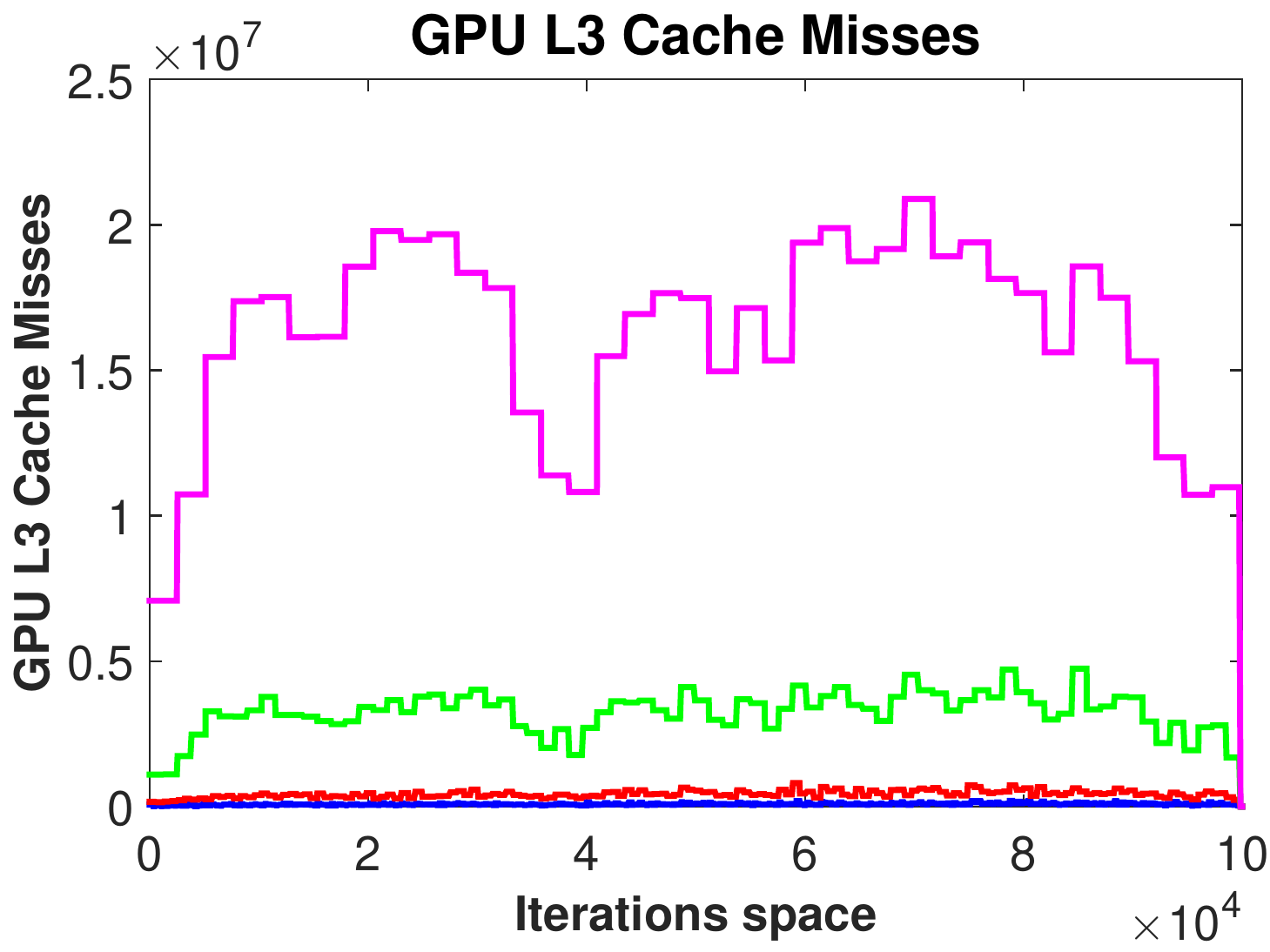}%
           \caption{}\label{fig:bhmisses}
         \end{subfigure}
       \begin{subfigure}[b]{0.33\textwidth}
           \includegraphics[width=\textwidth]{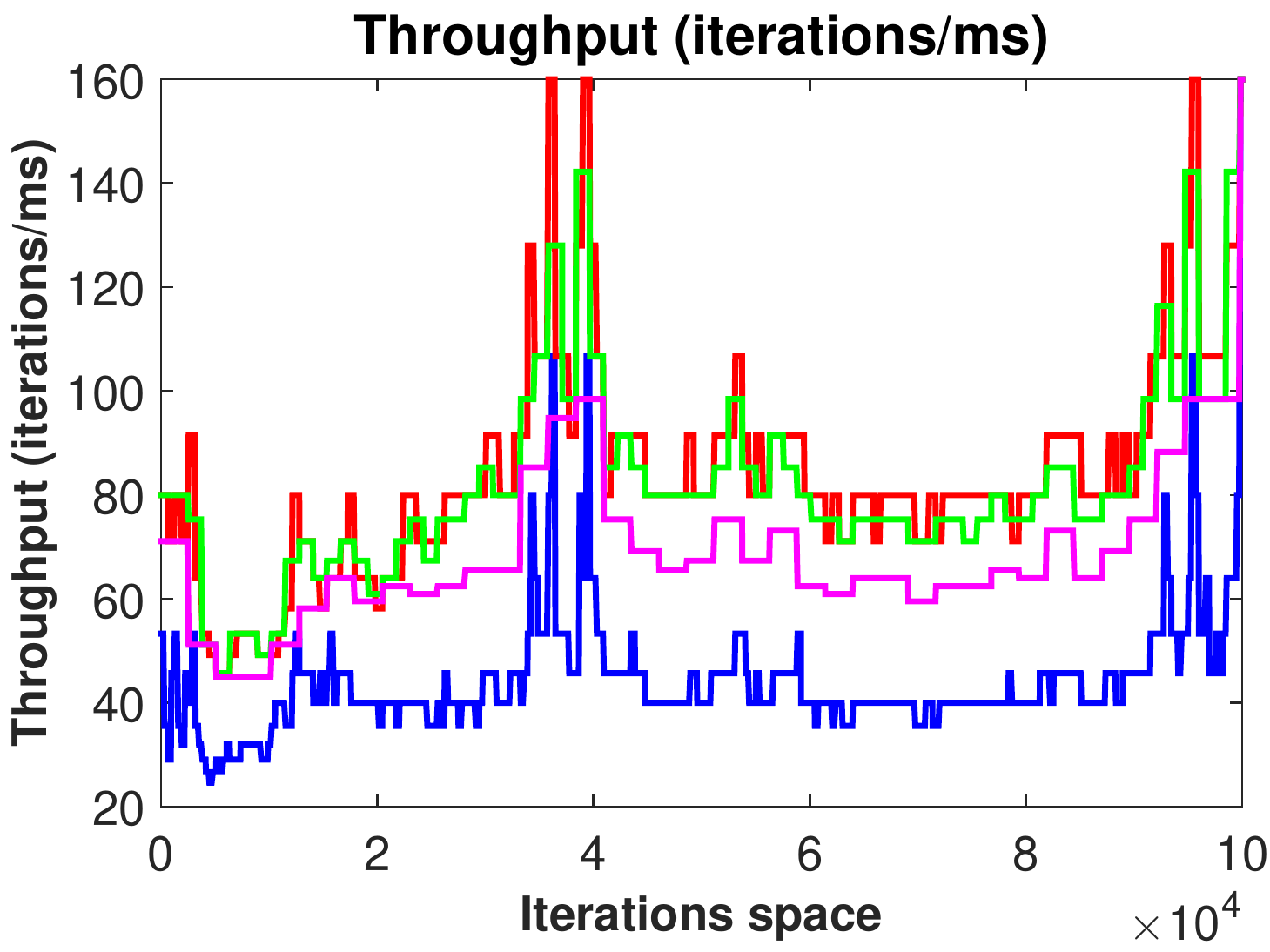}%
           \caption{}\label{fig:bhth}
        \end{subfigure}
     \caption{GPU Hardware metrics for Barnes Hut on Haswell. The
       legend of subfigure~\ref{fig:bhstalled} applies to all subfigures}\label{fig:BHmetrics}
\end{figure*}

The intensive use of the GPU offloading mechanism will reveal several
sources of overheads that have to be carefully considered. For
instance, not only the traditionally studied overheads like the data
transfers ones (host-to-device and device-to-host) have to be taken
into account, but also kernel launching and host thread dispatch
overheads gain relevance. In this paper, we study the impact of each
one of these overheads on different heterogenous architectures with an
integrated GPU. More precisely, in Section~\ref{sec:experimental}, we
conduct our experiments on two Intel processors (Ivy Bridge and
Haswell architectures) and one Samsung Exynos 5 featuring a big.LITTLE
architecture. Two different operating systems (Windows and Linux) are
considered.

With the information provided by these experiments we tackle the
problem of reducing the impact of the more representative
overheads. In Section~\ref{sec:optimizations}, we propose and evaluate
some optimizations that not only reduce the execution time but also
the energy consumption. Finally, we present the related works in
Section~\ref{sec:related} and conclude that in order to squeeze the
last drop of performance out of these heterogeneous chips, it is
mandatory to conduct a thorough analysis of the overheads and to
study how the CPU cores, the GPU and the different layers of
the software stack (OpenCL driver, OS, etc) interplay (see
Section~\ref{sec:conclusions}).

\section{Motivation}\label{sec:motivation}

We have found that, for irregular applications, offloading big chunk sizes
to the GPU can hinder performance. 
This is illustrated in Figure~\ref{fig:BHmetrics}, that shows the
evolution of different GPU hardware metrics on the Haswell
architecture (described in section~\ref{sec:experimentalsetting})
through the iteration space of one time step of the irregular
benchmark Barnes Hut. An input set of 100,000 bodies was used to
collect these results.  Each subfigure represents the evolution of the
metric of interest for different chunk sizes assigned to the GPU (see
chunk sizes legend in subfigure~\ref{fig:bhstalled}). We have used
Intel VTune Amplifier 2015~\cite{vtune15} to trace the ratio of cycles when EUs
(Execution Units) are active ($\sum_{all~EUs}$ cycles when EU
executes instructions/$\sum_{all~EUs}$ all cycles); the ratio of
cycles when EUs are idle ($\sum_{all~EUs}$ cycles when no threads
scheduled on EU/$\sum_{all~EUs}$ all cycles); the ratio
of cycles when EUs are stalled ($\sum_{all~EUs}$ cycles when EU
does not execute instructions and at least one thread is
scheduled on EU/$\sum_{all~EUs}$ all cycles); and the L3 cache
misses due to GPU memory requests. Subfigure~\ref{fig:bhth} also shows
the GPU effective throughput, measured as the number of iterations per
ms., through the iteration space. Data transfer and kernel offloading
overheads have been included in the computation of the throughput.

As we can see in subfigure~\ref{fig:bhth}, for the time step studied,
the optimal chunk size is 640 (see red line in the figure). Increasing
the chunk size beyond this value degrades the throughput. The hardware
metrics indicate that small chunk sizes (e.g.  320) do not effectively
fill the GPU computing units, as the ratio EU Idle indicates in
subfigure~\ref{fig:bhidle} (see blue line). However, when the chunk
size is large enough to fill the EUs (EU Idle <0.1), the EUs might
stall when the chunk size increases, due to the increment in L3 cache
misses (see green and pink lines in subfigure~\ref{fig:bhmisses}). This
is what happens in our irregular benchmark in which the majority of
memory accesses are uncoalesced. In our case, chunk sizes larger than
1280 dramatically increase L3 misses, which in turn increases the
ratio of EU Stalled (>0.9) and reduces the ratio of EU Active (<0.08),
causing a reduction of the effective throughput.

\begin{figure*}[htb]
        \centering
        \begin{subfigure}[b]{0.3\textwidth}
           \includegraphics[width=\textwidth]{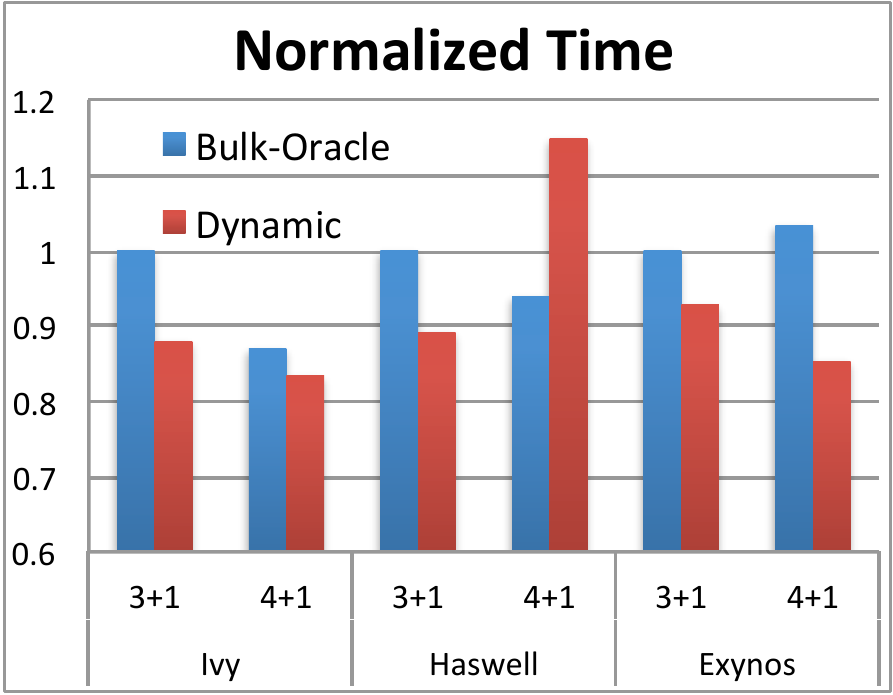}%
        \end{subfigure}
        ~~ 
        \begin{subfigure}[b]{0.3\textwidth}
           \includegraphics[width=\textwidth]{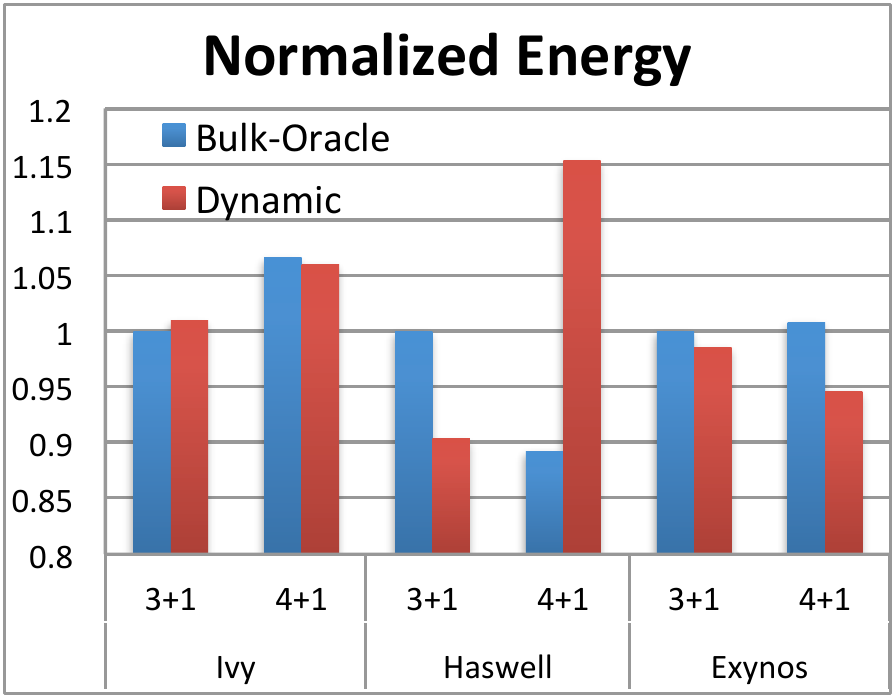}%
         \end{subfigure}
  ~~  \begin{subfigure}[b]{0.3\textwidth}
           \includegraphics[width=\textwidth]{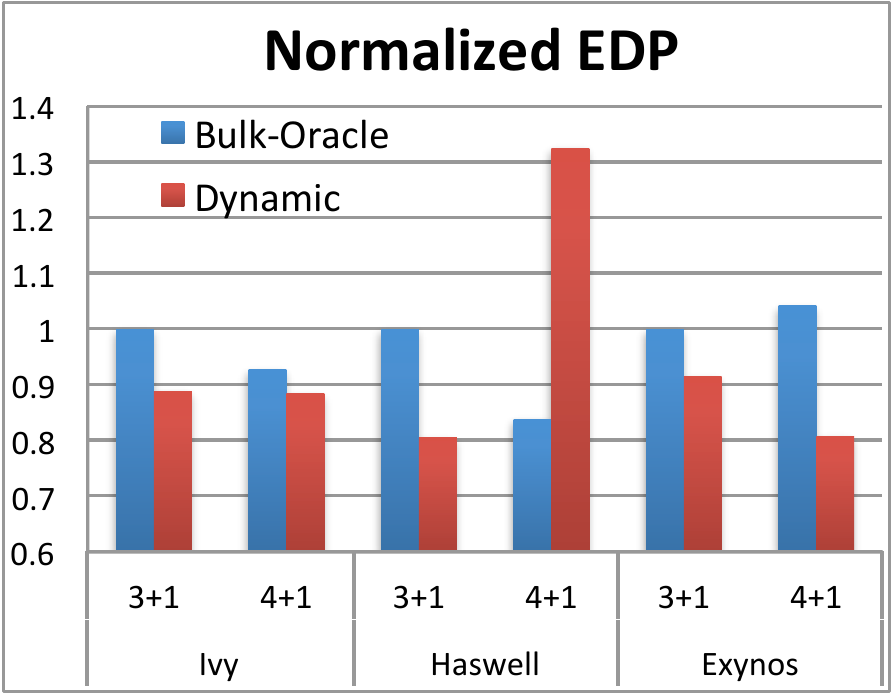}%
         \end{subfigure}
   \caption{Normalized time, energy and energy-delay-product (EDP) for
     Barnes Hut on Ivy Bridge, Haswell and Exynos. The lower the values
     the better.}\label{fig:BHTimeEnergyMotivation}
\end{figure*}

Therefore, in the quest of finding the optimal distribution of work
between the GPU and the CPU, if we assign a big chunk of iterations to
each device we can end up by not exploiting the GPU EUs
optimally. Thus, the partitioning strategy must be aware of the
optimal chunk of iterations that must be offloaded to the GPU for the
corresponding application. On the other hand, we have observed that
the effective throughput of the CPU cores is not so sensible to the
chunk size. As long as the chunk size is bigger than a threshold value,
the CPU throughput tends to be constant. For instance, when using the
Threading Building Blocks library (TBB)~\cite{oreilly:reinders:tbb},
it is recommended to have CPU chunk sizes that take 100,000 clock
cycles at least. By allowing a dynamic scheduling of the iteration
space in such a way that each device gets an optimal chunk of
iterations while balancing the workload among them, we can
minimize the execution time. However, some overheads are involved in
this type of scheduling, as we will see in the next sections. Anyway,
we explore here the applicability of a dynamic scheduling strategy on
a heterogeneous architecture using Barnes Hut as example.

In Figure~\ref{fig:BHTimeEnergyMotivation} we compare the time, energy
and EDP (Energy-Delay-Product)~\cite{Gonzalez96} for Barnes Hut on
three different heterogeneous architectures using 4 CPU cores and 1
integrated GPU: Ivy Bridge, Haswell and Exynos.  The input set also
has 100,000 bodies, but now 75 time-steps were computed. For each
platform, two scenarios are shown: \texttt{3+1}, which corresponds to
the case in which 4 threads are scheduled in the processor (3 CPU
threads + 1 host thread dedicated to offload the work to the GPU) and
\texttt{4+1} which corresponds to 5 threads (4 CPU threads + 1 host
thread). The later represents the case in which 1 thread of
oversubscription is allowed. We explore this case because we want to
make the most of our computing resources: since the host thread is
most of the time blocked while the GPU takes care of its chunk of
iterations, we add an additional thread to avoid one core becoming idle. 

In the figure, we call \texttt{Dynamic} to our dynamic
scheduling strategy (it works similarly to the OpenMP dynamic
scheduling policy of the \texttt{pragma omp parallel for}). In this strategy, first we have
to find the optimal chunk size for the GPU (the GPU chunk). This is
done through an offline training phase, where we explore different
chunk sizes, and choose the value that maximizes the effective
throughput of the application in the corresponding
GPU. Table~\ref{tab:BHchunksize} shows the optimal size for each
platform. This size is passed to our scheduler which dynamically
assigns a new chunk to the GPU each time it finishes the
computation. Similarly, each CPU core gets a new chunk of iterations
(the CPU chunk) every time it finishes the previous one. The size of
the CPU chunk is selected to balance the load with the GPU computation
(details are provided in the next section).

We compare this dynamic strategy with a static scheduling that assigns
a single GPU chunk to the GPU and the rest of the iterations to the
CPU cores. Previously, for the static scheduling we carry out an
exhaustive offline profiling that looks for the static partitioning of
the iteration space between the CPU and GPU that minimizes the
execution time. We vary the percentage of the iteration space
offloaded to the GPU (between 0\% -only CPU execution- and 100\% -only
GPU execution- using 10\% steps). We call this strategy
\texttt{Bulk-Oracle}. Again, Table~\ref{tab:BHchunksize} shows the
optimal percentage of iterations offloaded to the GPU for each
platform under the static partition strategy. Notice that with the
static approach the whole chunk is offloaded to the GPU at the
beginning of each time step.

 \begin{table}[!hbt]
\centering
\caption{Optimal GPU chunk size for Dynamic and optimal percentage of
  the iteration space offloaded to the GPU for Bulk-Oracle}\label{tab:BHchunksize}
 \begin{tabular}{l|cc|cc|cc|}\cline{2-7}
        & \multicolumn{2}{|c|}{Ivy} &  \multicolumn{2}{|c|}{Haswell} &\multicolumn{2}{|c|}{Exynos} \\\cline{2-7}
     & 3+1 & 4+1 & 3+1 & 4+1 & 3+1 & 4+1 \\ \cline{2-7}
Dynamic &1536 & 1536& 2048& 2048 & 2048& 2048\\\cline{2-7}
Bulk-Oracle &50\% &40\% &70\% &70\% & 20\%&20\% \\\cline{2-7}
 \end{tabular}
 \end{table}

 In Figure~\ref{fig:BHTimeEnergyMotivation} each parameter (time,
 energy and EDP) has been normalized with respect to the value
 obtained for the Bulk-Oracle 3+1 execution on each architecture. As
 we see, the dynamic strategy outperforms the static one
 (Bulk-Oracle), except in the case of Haswell for 4+1, where overheads
 associated to the host thread degrade performance (both in time and
 energy). This issue is discussed in the next section. Another
 interesting result is that oversubscription (4+1) improves the
 execution times on Ivy, for both the static and dynamic approaches,
 but it also increases the energy consumption. This is due to the fact
 that 4 threads compute a larger number of chunks on the CPU cores
 and, since the CPU is less energetically efficient computing the
 chunks than the GPU, this results in higher energy consumption. For
 this architecture and this benchmark, the increment of the energy on
 the CPU is not amortized by the reduction of time. On the other hand,
 on Exynos, a static partitioning does not scale when going from 3+1
 to 4+1, while the dynamic strategy reduces time and energy
 consumption.

\section{Scheduling Strategy}\label{sec:scheduling}

\vspace{0.3in}

In this section, we present in more detail our scheduler
(Section~\ref{sec:scheduler}), how the partitioner works
(Section~\ref{sec:partition}), and the potencial
sources of overhead (Section~\ref{overheads}).

\subsection{Scheduler description} 
\label{sec:scheduler}

Our scheduler, that we call Dynamic, considers loops with independent
iterations (\texttt{parallel\_for}) and features a work scheduling
policy with a dynamic GPU and CPU chunk partitioning. Our approach
dynamically partitions the whole iteration space into chunks or blocks
of iterations. The goal of the partitioning strategy is to evenly
balance the workload of the loop among the compute resources (GPU and
CPU cores) as well as to assign to each device the chunk size that
maximizes its throughput. This is key because, as shown in
Figure~\ref{fig:BHmetrics}, the chunk size can have a significant
impact on the performance of heterogeneous architectures, especially
when dealing with irregular codes.


\begin{figure}
\centering
\begin{minipage}{0.9\linewidth}
\begin{lstlisting}[style=C]
#include <HScheduler.h>
class Body{(*@\label{list:bodyObjectstart}@*)
public:
  void operatorCPU(int begin, int end) { (*@\label{list:operatorCPUstart}@*)
     for(i=begin; i!=end; i++){ ... }
  }(*@\label{list:operatorCPUend}@*)
  void operatorGPU() (int begin, int end){(*@\label{list:operatorGPUstart}@*) 
     hostToDevide(begin, end){...}(*@\label{list:operatorGPUh2d}@*)
     launchKernel(begin, end){...}(*@\label{list:operatorGPUl}@*)
     deviceToHost(begin, end){...}(*@\label{list:operatorGPUd2h}@*)(*@\label{list:operatorGPUend}@*)
     clFinish();
  }
}(*@\label{list:bodyObjectend}@*)
...
int main(int argc, char* argv[]){
  
  Body body;

  // Start task scheduler
  task_scheduler_init init(nThreads);(*@\label{list:init}@*)
  ...
  parallel_for(begin,end,body,Partitioner_H(G));(*@\label{list:for}@*)
  ...
}
\end{lstlisting}
\end{minipage}
\caption{Using the \texttt{parallel\_for} template}
\label{fig:parallelforcode}
\end{figure} 

Our scheduler builds on top of an extension of the TBB
\texttt{parallel\_for} template for heterogenous GPU-CPU systems by
Navarro et al.~\cite{TRNavarro}. Figure~\ref{fig:parallelforcode}
shows the pseudo-code to use the extended \texttt{parallel\_for}
construct in such heterogeneous systems. As in any TBB program, the
scheduler is initialized (line \ref{list:init}). In this step, the
developer sets the number of OS threads, \texttt{nThreads}, that the
TBB runtime will create, which can vary from 1 to the number of CPU
cores plus one additional thread to host the GPU (the host
thread). Then, the developer can invoke the \texttt{parallel\_for}
(line \ref{list:for}), which has the following parameters: the
iteration space (the range \texttt{begin, end}), the body object of
the loop (\texttt{body}), and the partitioner object
(\texttt{Partitoner\_H}).  The latter argument, effectively overloads
the native TBB \texttt{parallel\_for} function so that the
heterogeneous version is invoked. Besides, the \texttt{Partitoner\_H}
method takes care of the dynamic partitioning strategy that gets the
optimal chunk size for the GPU (parameter \texttt{G} provided by the
user) and for the CPU cores as described in
Section~\ref{sec:partition}.

The user also writes the \texttt{class Body} that
processes the chunk on the CPU cores or on the GPU 
(lines~\ref{list:bodyObjectstart}-\ref{list:bodyObjectend} in
Figure~\ref{fig:parallelforcode}). Two methods (operators) must be
coded. One for the CPU 
(lines~\ref{list:operatorCPUstart}-\ref{list:operatorCPUend}) and one
for the GPU
(lines~\ref{list:operatorGPUstart}-\ref{list:operatorGPUend}).  For
the GPU, the user has to define two functions to perform the
asynchronous host-to-device (line~\ref{list:operatorGPUh2d}) and
device-to-host (line~\ref{list:operatorGPUd2h}) memory transfers, as
well as the kernel launching (line~\ref{list:operatorGPUl}). Since
these functions are all asynchronous, we finish the GPU part of the
body with the synchronous \texttt{clFinish()} call that does not
return until all the previous steps have been completed.

Internally, our scheduler is implemented as a pipeline that consists
of two stages or filters: Filter$_1$, which selects the
computing device 
and the chunk size (number of iterations) assigned, and Filter$_2$,
which processes the chunk on the corresponding device. Filter$_1$
firstly checks if the GPU device is available. In that case, a
G\_token is created and initialized with the range of the GPU chunk.
If there is no idle GPU device, then a CPU core is idle; thus, a
C\_token is created and initialized with the range of the CPU
chunk. In both cases, the partitioner extracts a chunk of iterations from
the range of the remaining iteration space.  Next, Filter$_2$ processes the
chunk in the corresponding device
and records the time it takes to compute 
the corresponding chunk\footnote{Note that for the GPU, the time includes the
  kernel execution time as well as the hostToDevice and deviceToHost times.}. 
This is necessary to compute the device's throughput, which
is used by the partitioner described next.

\subsection{Partitioning strategy} 
\label{sec:partition}

We assume that the execution time can be seen as a sequence of
scheduling intervals $\{tG_0, tG_1 \ldots tG_{i-1}, tG_i, tG_{i+1}
\ldots\}$ for the GPU and $\{tC_0, tC_1 \ldots tC_{i-1}, tC_i,
tC_{i+1} \ldots\}$ for each CPU core. Each computing device at the
current interval, $tG_i$ or $tC_i$, can get a chunk of iterations. The
running time $T(tG_i)$, for each GPU's chunk size $G(tG_i)=G$, or
the time $T(tC_i)$ for a CPU's chunk size $C(tC_i)$, is
recorded. This time is used to compute the throughput,
$\lambda_G(tG_i)$ for the GPU or $\lambda_C(tC_i)$ for a CPU core, in
the current scheduling interval as,

\vspace*{-3mm}
{\small
\begin{eqnarray}
  \label{eq:lambG}
\lambda_G(tG_i)&=& \frac{G}{T(tG_i)} \\
  \label{eq:lambC}
\lambda_C(tC_i)&=& \frac{C(tC_i)}{T(tC_i)}  
\end{eqnarray}
}
\vspace*{-3mm}


In order to compute the chunk size for the GPU, an offline training
phase explores different chunk sizes, and chooses the value that
maximizes the effective throughput of the application for a given
input data.  To reduce
the number of runs of this offline training phase, we set the GPU
chunk size to the smallest number of iterations that fully occupy the
GPU resources. For example, on the integrated GPU of the Intel Haswell
we have 20 EU (execution units) each one running a SIMD-thread (aka
wavefront) at a time, and 8, 16 or 32 work-items per SIMD-thread
(decided at compile time). In
OpenCL these values can be queried reading two
variables\footnote{\small CL\_DEVICE\_MAX\_COMPUTE\_UNITS,
  and\\ CL\_KERNEL\_PREFERRED\_WORK\_GROUP\_SIZE\_MULTIPLE}.  Thus,
the product of these two arguments is used as the initial GPU chunk
size. For Barnes Hut the compiler chooses 16 SIMD and therefore we
select an initial GPU chunk size of $20\times 16=320$
iterations. Then, chunk sizes that are multiple of this initial chunk
size are tried.  We keep trying different chunk sizes while the
throughput increases.  Once the throughput decreases or remains stable
for 2 or more chunk sizes, the training phase ends and the GPU chunk
size that obtained the highest throughput is chosen. This approach
requires only a few runs.

Regarding the policy to set the CPU core chunk size, our partitioner
follows the heuristic by Navarro et al.~\cite{TRNavarro}. Basically,
this heuristic is based on a strategy to minimize the load imbalance
among the CPU cores and the GPU, while maximizing the throughput of
the system. To that end, the optimization model described there
recommends that: {\it each time that a chunk is partitioned and
  assigned to a device, its size should be selected such that it is
  proportional to the device's effective throughput}. Therefore, we
implemented a greedy partitioning algorithm based on the following key
observation: while there are sufficient remaining iterations, the
chunk size assigned to a GPU at the scheduling interval $tG_i$
should be the optimal for the GPU ($G$, as explained in the previous
paragraph), whereas at the scheduling interval $tC_i$ the chunk size
assigned to a CPU core, $C(tC_i)$, should verify:


{\small
\begin{equation}
\frac{C(tC_i)}{\lambda_{C}(tC_{i-1})} = \frac{G}{\lambda_G(tG_{i-1})}
\label{eq:der1}
\end{equation}
}
\vspace*{-3mm}

\noindent where $\lambda_{C}(tC_{i-1})$ and $\lambda_{G}(tG_{i-1})$ are
the CPU and GPU throughputs in the previous 
scheduling intervals, respectively. So we have:

\vspace*{-3mm}
{\small
\begin{equation}
C(t_i) = G\cdot \frac{\lambda_{C}(tC_{i-1})}{\lambda_{G}(tG_{i-1})}
\label{eq:der2}
\end{equation}
}
\vspace*{-3mm}

\subsection{Sources of overhead}
\label{overheads}

\begin{figure}[htb]
    \centering
    \includegraphics[width=0.7\linewidth]{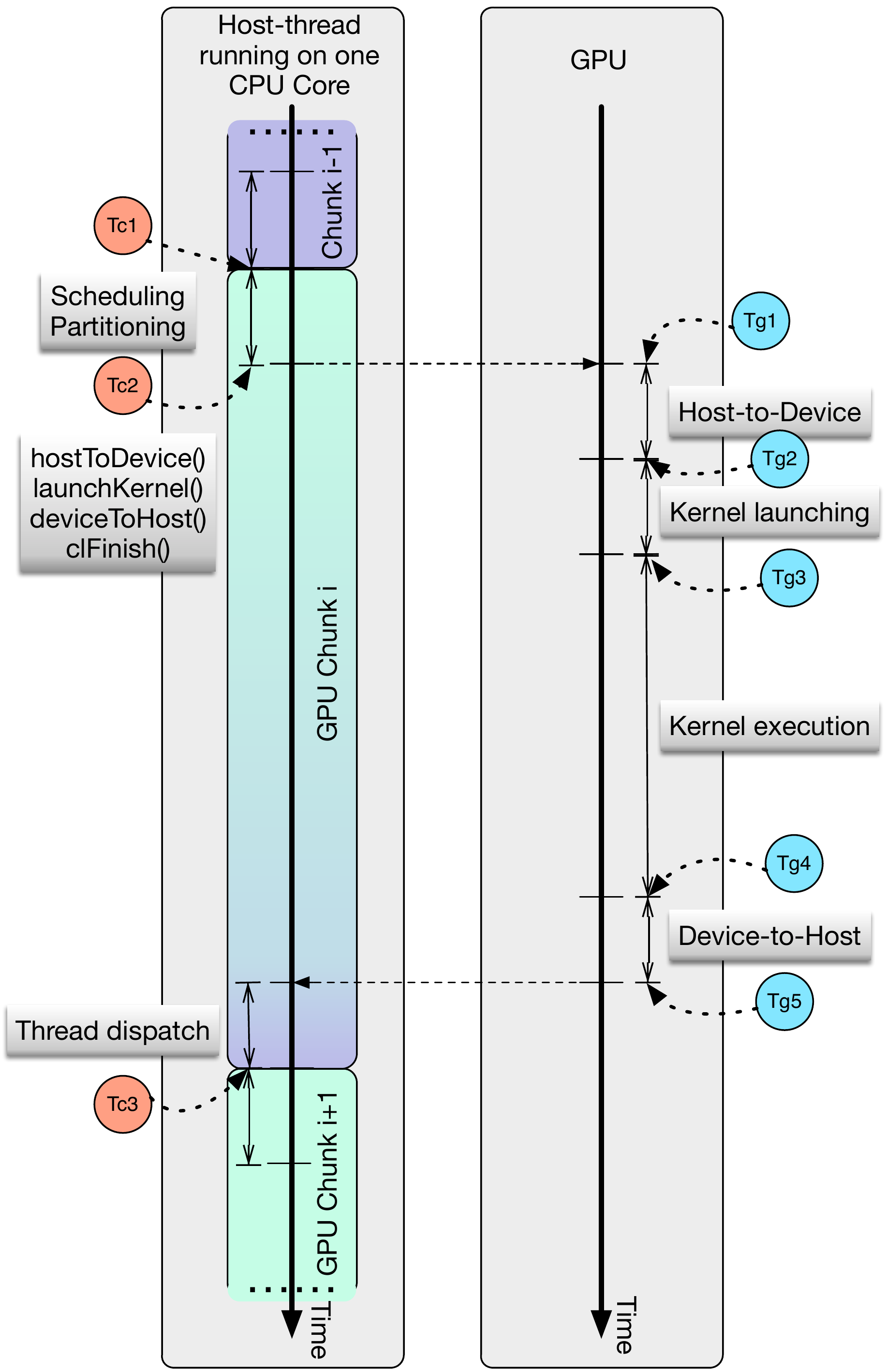}%
    \caption{Time diagram showing the different phases of offloading a
    chunk to the GPU}\label{fig:typeofoverheads}
\end{figure}

Figure~\ref{fig:typeofoverheads} shows the different phases that our Dynamic 
framework has to perform to offload a chunk of iterations to the GPU. 

As we explained in the previous section, our framework uses nThreads,
from which one of them is called the host thread and just takes care
of serving the GPU. This host thread runs in one of the available
cores and first executes the code associated with the scheduler and
the partitioner (the code of Filter$_1$ explained earlier). Then, the same
host thread executes the Filter$_2$ stage that includes the function
calls listed in Figure~\ref{fig:parallelforcode}
(\texttt{hostToDevice()}, \texttt{launchKernel()},
\texttt{deviceToHost}, and \texttt{clFinish()}). The first three calls
just asynchronously enqueue the corresponding operation on the GPU's
command queue,
whereas the latter is a synchronous wait. In Figure~\ref{fig:typeofoverheads}
we can see that the enqueued operations are sequentially executed on
the GPU where we consider the times taken by the ``Host-to-Device'',
``Kernel launching'', ``Kernel execution'' and
``Device-to-Host''. When this last operation is done, the host thread
is notified but some time may be taken by the OS to re-schedule the
host thread. This time is illustrated in the figure with the label ``Thread
dispatch''.

\begin{figure*}[htb]
        \centering
        \begin{subfigure}[b]{0.3\textwidth}
           \includegraphics[width=\textwidth]{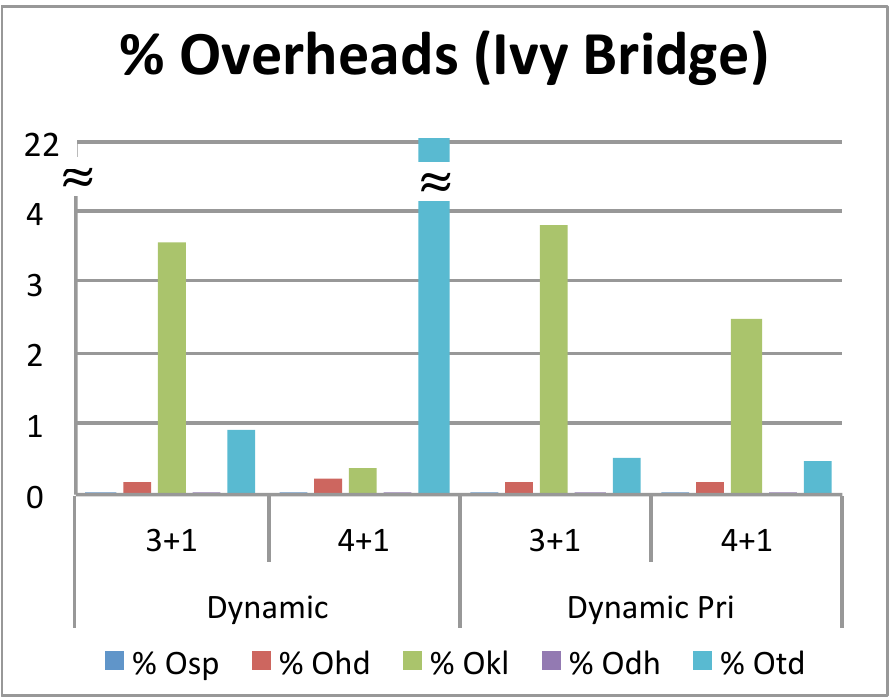}%
        \end{subfigure}
        ~~ 
        \begin{subfigure}[b]{0.3\textwidth}
           \includegraphics[width=\textwidth]{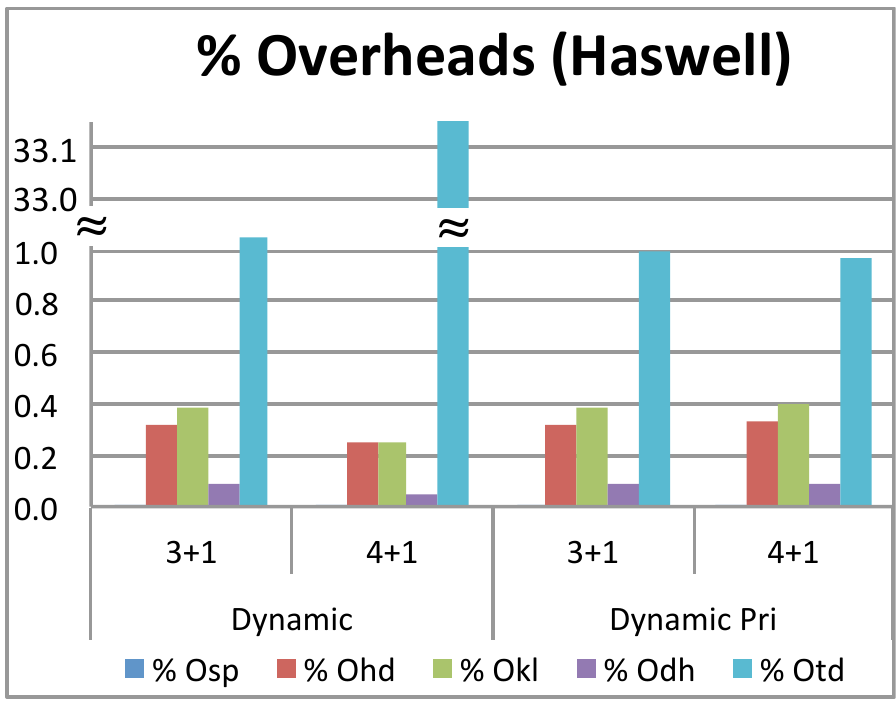}%
         \end{subfigure}
  ~~  \begin{subfigure}[b]{0.3\textwidth}
           \includegraphics[width=\textwidth]{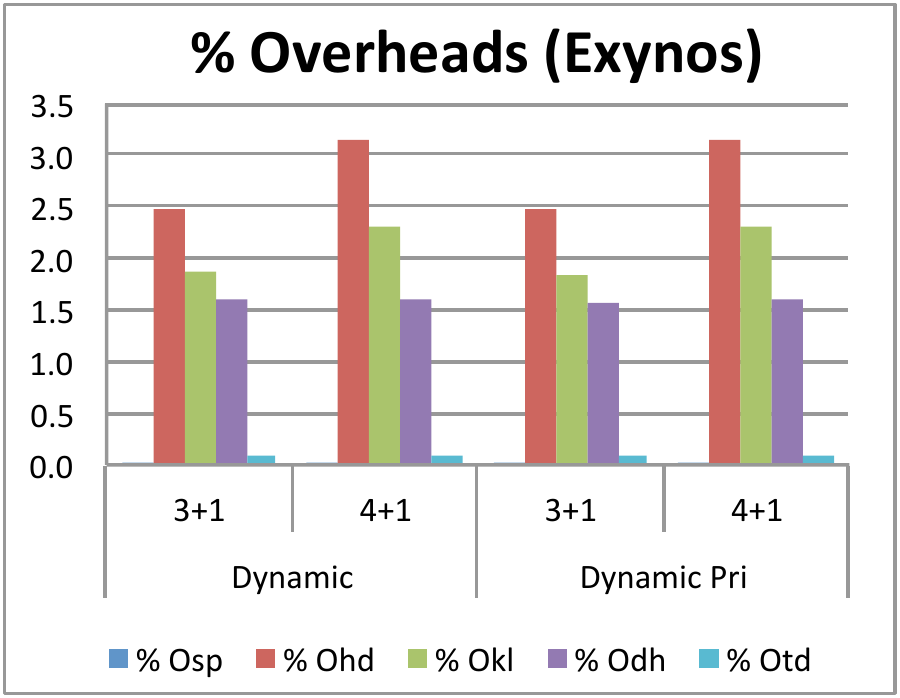}%
         \end{subfigure}
   \caption{Percentage of the different overheads for Barnes Hut on
Ivy Bridge,  Haswell and Exynos. The lower, the better.}\label{fig:overheads}
\end{figure*}

In order to measure the relevant overheads involved in the execution
of the code some time stamps are taken on the CPU
(Tc1, Tc2 and Tc3) and on the GPU (Tg1 to Tg5) as depicted in
Figure~\ref{fig:overheads}. To get the CPU time stamps we rely on TBB's
\texttt{tick\_count} class, whereas for the GPU we configure the OpenCL
command queue in the profile mode so that we can read the ``start'' and
``complete'' time stamps of each of the enqueued commands.

With this information we compute the overhead of Scheduling and
Partitioning, $O_{sp}$, Host-to-Device operation, $O_{hd}$, Kernel
Launching, $O_{kl}$, Device-to-Host, $O_{dh}$, and Thread Dispatch,
$O_{td}$, as follows:

 \vspace*{-3mm}
{\small
\begin{eqnarray}
  \label{eq:osp}
O_{sp}&=& \frac{\sum_{\#GPUchunks}{(Tc2-Tc1)}}{Total Execution Time} \\
  \label{eq:ohd}
O_{hd}&=& \frac{\sum_{\#GPUchunks}{(Tg2-Tg1)}}{Total Execution Time} \\
  \label{eq:okl}
O_{kl}&=& \frac{\sum_{\#GPUchunks}{(Tg3-Tg2)}}{Total Execution Time} \\
  \label{eq:odh}
O_{dh}&=& \frac{\sum_{\#GPUchunks}{(Tg5-Tg4)}}{Total Execution Time} \\
  \label{eq:otd}
O_{td}&=& \frac{\sum_{\#GPUchunks}{\Bigl((Tc3-Tc2)-(Tg5-Tg1)\Bigr)}}{Total Execution Time} 
\end{eqnarray}
}
\vspace*{-3mm}

\section{Analysis of Overheads}\label{sec:experimental}

In this section we analyze the sub-optimal performance of the Dynamic
scheduler for the 4+1 configuration, as shown in
Figure~\ref{fig:BHTimeEnergyMotivation}. To better understand the
underlying reasons for that performance degration we first describe
our environment setup (Section~\ref{sec:experimentalsetting})
and then discuss the overheads that we measure
(Section~\ref{discussion}).

\subsection{Experimental Settings}\label{sec:experimentalsetting}
We run our experiments on three different platforms: two Intel-based
desktops and an Odroid XU3 bare-board. More precisely, the desktops
are based on two quad-core Intel processors: a Core i5-3450, 3.1GHz,
based on the Ivy Bridge architecture, and a Core i7-4770, 3.4GHz,
based on Haswell. Both processors feature an on-chip GPU, the HD-2500
and HD-4600, respectively.  The Odroid has a Samsung Exynos 5 (5422)
featuring a Cortex-A15 2.0Ghz quad core along with a Cortex-A7 quad
core CPUs (ARM's big.LITTLE architecture). This platform features
current/power monitors based on TI INA231 chips that enable readouts
of instant power consumed on the A15 cores, A7 cores, main memory and
GPU.  The Exynos 5 includes the GPU Mali-T628 MP6.

Regarding the software tools, on the desktops we rely on Intel
Performance Counter Monitor (PCM) tool~\cite{pcm} to access the HW
counters (which also provides energy consumption in Joules). The GPU
kernels are implemented with Intel OpenCL SDK 2014 that is currently
only available for Windows OS. The benchmarks are compiled using Intel
C++ Compiler 14.0 with -O3 optimization flag. The
Odroid board runs Linux Ubuntu 14.04.1 LTS, and an in-house library
has been developed to measure energy consumption, as
described below.  On this platform, the compiler used is gcc 4.8 with -O3 flag. 
The OpenCL SDK v1.1.0 is used for the Mali GPU. 

On all platforms, Intel TBB 4.2 provides the core template of the heterogenous
\texttt{parallel\_for}.  We measured time and energy in 10 runs of the
applications and report the average.

\subsubsection{Energy measurement on the Exynos 5}\label{sec:meter}

The Odroid XU3 platform is shipped with an integrated power analysis
tool. It features four on-board INA231 current/power monitors\footnote{http://www.ti.com/product/ina231} to monitor the Cortex A15 cores, Cortex A7 cores, DRAM and GPU power dissipation.
The readouts from these power sensors are accessible through the /sys
file system from user-space. The system does not provide cumulated
energy consumptions, so only instant power readings are available.

We have developed a library to measure the energy consumed 
by our executions. The library allows starting/stopping a dedicated
thread to sample power readings and integrate them through time using
the real-time system clock. Once the sampling thread is working, the
library also provides functions to take partial energy measurements to
profile the energy consumption through different algorithm stages. 
It is possible to monitor energy (in Joules) consumed by the Cortex 
A15 cores, the Cortex A7 cores, DRAM and GPU, separately. 
The energy figures on the Exynos 5 we present
in this paper are the sum of the four power monitors aforementioned.

The sample rate used is 10 Hz. Thus, one sampled power value is
obtained every 100 milliseconds. The power value read is multiplied by
the sampling period and the product is integrated in a cumulated
energy value (in Joules). We have chosen this sample rate because the
power sensors actualize their values every 260 milliseconds
approximately, so a sample rate two times faster is good enough for
getting accurate measurements (sampling rate is below Nyquist
rate).



\subsection{Discussion}
\label{discussion}

Figure~\ref{fig:overheads} shows the overheads that we measured for
the corresponding terms defined in the previous section
(eqs.~\ref{eq:ohd}-\ref{eq:otd}) for the three platforms that we
consider, Ivy Bridge, Haswell and Exynos when running the Barnes Hut
benchmark with 100,000 bodies and 15 time steps for the 3+1 and 4+1
scenarios. This section covers only the Dynamic results (shown on the
left side) of the three subfigures in Figure~\ref{fig:overheads}.

The smallest overhead, on average, is the one due to Scheduling and
Partitioning, $O_{sp}$, that represents 0.02\% on Ivy Bridge, and less than
0.004\% on Haswell and Exynos. 

Regarding the data transfer overheads, $O_{hd}$ and $O_{dh}$, on Ivy
Bridge and Haswell, they are always below 0.3\%. However, on Exynos,
these overheads are significantly larger, around an order of magnitude
higher: on average $O_{hd}=$2.8\% and $O_{dh}$=1.6\%. After testing
our platforms with memory bound micro-benchmarks, we found that Exynos
exhibits an order of magnitude higher data transfers times than the
Intel architectures. This explains the impact of Host-To-Device and
Device-To-Host overheads on the Exynos. More precisely, the
\texttt{hostToDevice()} operation implicitly copies the host buffer onto
a different region of memory that can be accessed by the GPU and that
is non-pageable (pinned memory). Similarly, \texttt{deviceToHost()}
does another memory-to-memory copy operation in the other
direction. Therefore, lower memory bandwidth on the Exynos results in
more apreciable Host-To-Device and Device-To-Host overheads with
respect to the Intel architectures. As future work we will study how
Barnes Hut memory accesses could be reorganized so that the cores and
the GPU could share the same buffer, avoiding the copy
operations using the zero-copy-buffer capability of OpenCL.

\begin{figure*}[htb]
        \centering
        \begin{subfigure}[b]{0.3\textwidth}
           \includegraphics[width=\textwidth]{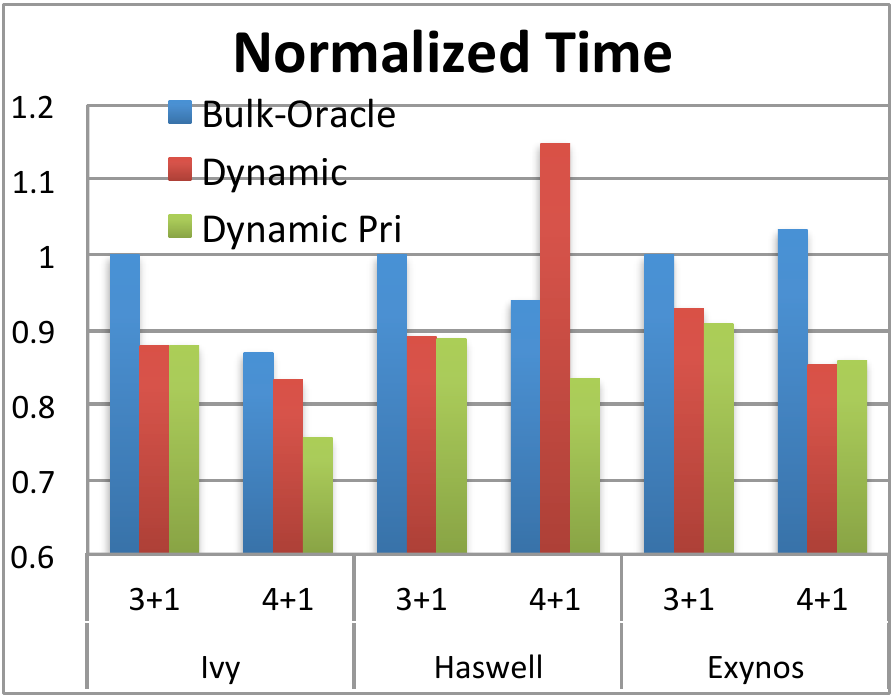}%
        \end{subfigure}
        ~~ 
        \begin{subfigure}[b]{0.3\textwidth}
           \includegraphics[width=\textwidth]{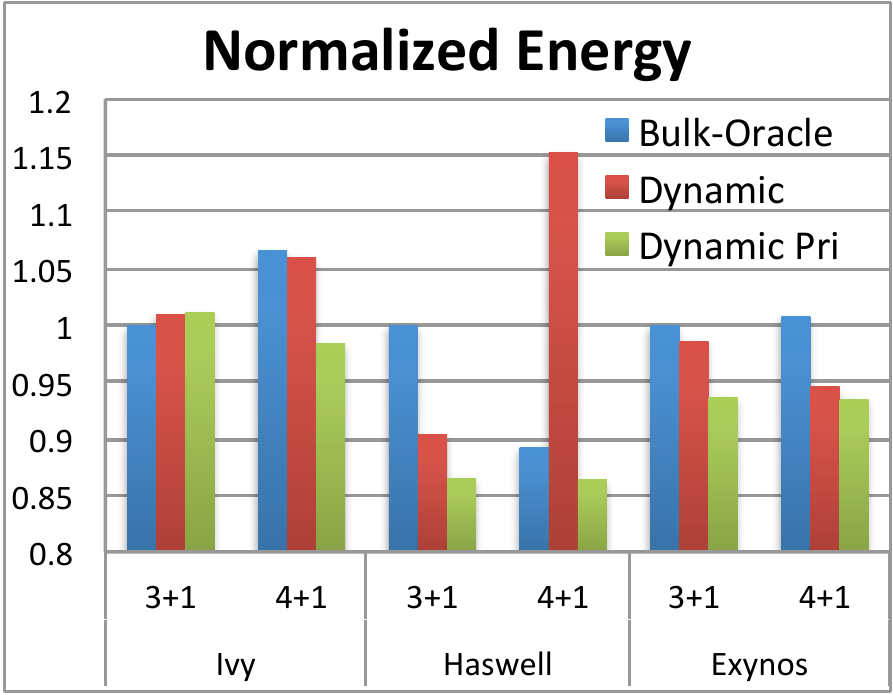}%
         \end{subfigure}
  ~~  \begin{subfigure}[b]{0.3\textwidth}
           \includegraphics[width=\textwidth]{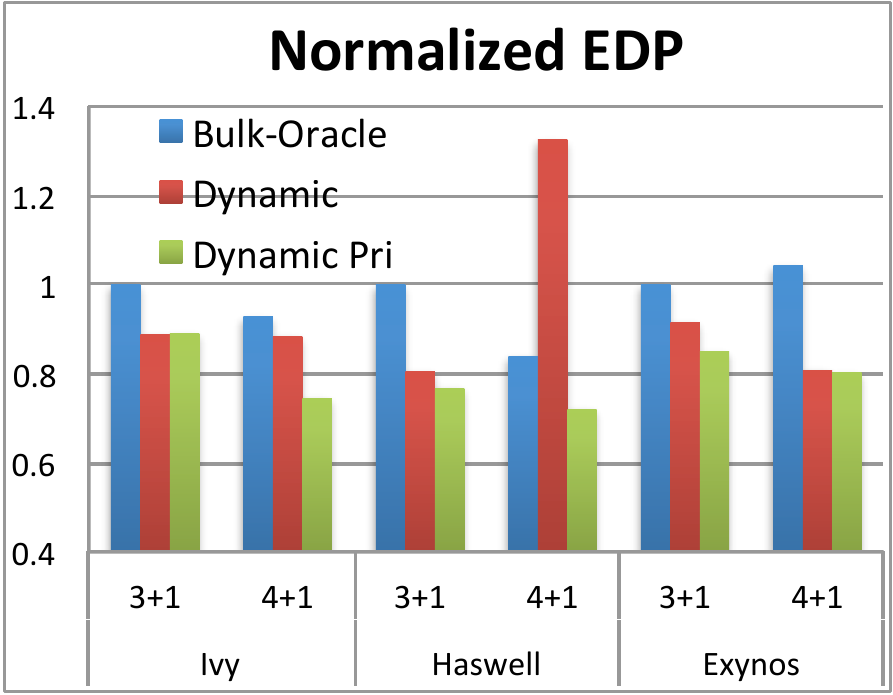}%
         \end{subfigure}
   \caption{Normalized time, energy and energy-delay-product (EDP) for
     Barnes Hut on Ivy Bridge, Haswell and Exynos adding a higher
     priority to the host-thread. The lower, the better.}\label{fig:BHTimeEnergyPri}
\end{figure*}

With respect to the Kernel Launch overhead, it goes from up to 0.4\%
on Haswell to up to 3\% and 2\% on Ivy Bridge and Exynos,
respectively. Average times consumed on this operation are 1.8 ms, 1 ms, 
and 3.6 ms on Ivy Bridge, Haswell and Exynos, respectively.

However, on the desktop platforms one of the most noticeable overheads
is $O_{td}$ (Thread Dispatch) especially for the 4+1 scenario. For
that case, it represents 22\% and 33\% of the total execution time on
Ivy Bridge and Haswell, respectively. Notice that this happens only
for the oversubscription cases and under Windows OS. Actually, on
Exynos under the Linux OS, the Tread Dispatch overhead represents less
than 0.09\% in all cases. The overhead in Windows is explained because
the OpenCL driver ends up blocking the host thread once it has
offloaded the kernel. It does that to avoid wasting a core with a
busy-waiting host thread. In the 4+1 scenario, 4 threads are already
intensively using the 4 cores and when the OpenCL driver notifies the
OS about the GPU completion, the OS wakes up the host thread. However,
if this host thread does not have enough priority it is unlikely it
will be dispatched to the running state straightaway. The core of the
Windows scheduling policy is Round Robin, so the just awaken host
thread has to wait on the ready queue to the next available time
slice. This does not happen in the Linux scheduler, where a just
awaken thread gains more priority than the other CPU-bound threads
that are intensively computing CPU chunks. This scheduling decision in
Linux rewards interactive threads (IO-bound). As a consequence, in
Linux, the host thread will be immediately dispatched after being
woken up. This result is corroborated thanks to the experiment
described in the next section.

\section{Optimizations}\label{sec:optimizations}

After identifying the main sources of overhead of our Dynamic approach, 
in this section we discuss the optimizations that can be implemented
to address them. The overarching goal is not only reduce the impact of the
overhead, but also to reduce the energy consumption. To that end, we
propose one strategy for the Windows based platforms (Section~\ref{priority}) and a different
one for the Exynos architecture (Section~\ref{big:little}).

\subsection{Increasing the priority of the host thread}
\label{priority}
As it has been shown in the previous section, on the Windows based
desktops, the highest overhead appears for the 4+1 configuration,
where we have measured that up to 22\% and 33\% of the execution time
is wasted on the \texttt{clFinish()} operation on the Ivy Bridge and
Haswell, respectively. This can be solved by assigning a higher
priority to the host thread so that another thread can be immediately
preempted and the just awoken host thread can take up a core and start
feeding the GPU again. This is key when the GPU processes the chunks
more efficiently than the CPU, as it happens in our benchmark. To
boost the host thread priority we rely on the
\texttt{SetThreadPriority()} Windows API.  The framework obtained with
this optimization is called Dynamic Pri.

The right part of the three subfigures of Figure~\ref{fig:overheads}
shows the overheads incurred by Barnes Hut when using Dynamic Pri and
the arguments described in the previous section. The figure shows that
on Ivy Bridge and Haswell, the $O_{td}$ overhead for the 4+1
configuration has been reduced as is now similar to that of the 3+1
one. This confirms that increasing the priority of the host thread for
the 4+1 configuration reduces this particular overhead. Note that on
the Exynos platform increasing the host thread priority barely affects
the measured overheads.

Figure~\ref{fig:BHTimeEnergyPri} shows the normalized metrics (time,
energy, and EDP) w.r.t. Bulk-Oracle 3+1, as shown in
Figure~\ref{fig:BHTimeEnergyMotivation}, but now adding a new bar
representing the results for Dynamic Pri.  As expected,
Figure~\ref{fig:BHTimeEnergyPri} confirms that for Ivy and Haswell,
that run Windows OS, boosting the priority of the host thread has
almost no impact in reducing time, energy or EDP for the 3+1
configuration. However, with respect to Dynamic, Dynamic Pri has a
significant impact on these metrics in the oversubscribed 4+1
scenario. For instance, on the Ivy Bridge, the Dynamic Pri reduces
time, energy, and EDP by 10\%, 7\% and 18\%, respectively, when
comparing with Dynamic. On Haswell, these reductions are even more
significant: 37\%, 33\% and 84\% for time, energy and EDP,
respectively. One interesting result appears on Ivy Bridge when
comparing the Dynamic 3+1 configuration versus the Dynamic 4+1 as we
mentioned in Section~\ref{sec:motivation}. Let's remind that even
though the 4+1 configuration is faster than the 3+1, the energy
consumed by the 4+1 is higher, which is not true any more for Dynamic
Pri. Now, Dynamic Pri uses the GPU more efficiently (it is served
quicker and this results in the GPU processing more
chunks). Therefore, the more energy consuming CPU cores end up doing
less work (and consuming less energy) resulting in smaller total energy
consumption figures.

\begin{figure*}[hbt]
        \centering
        \begin{subfigure}[b]{0.3\textwidth}
           \includegraphics[width=\textwidth]{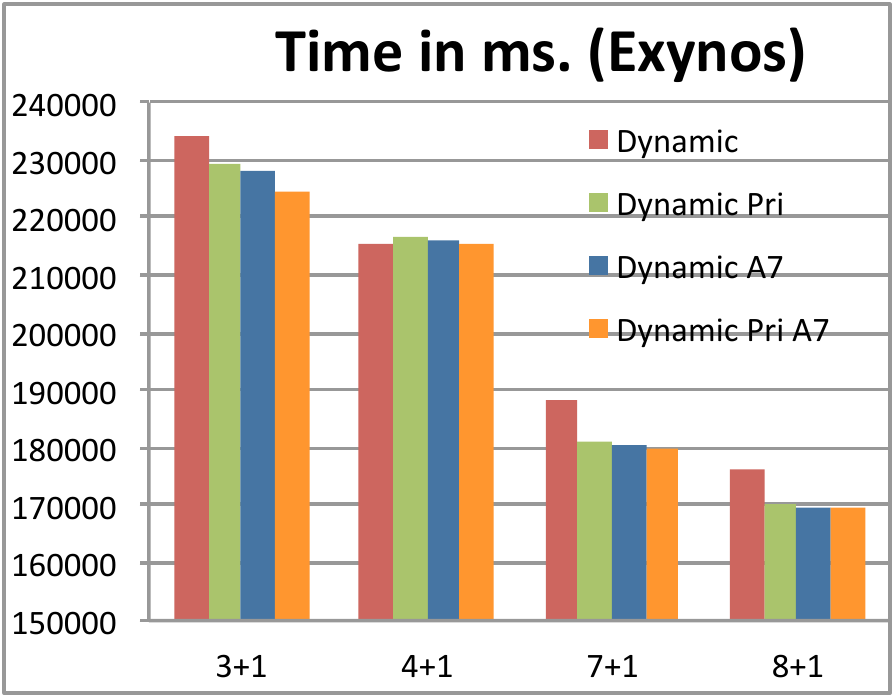}%
        \end{subfigure}
        ~~ 
        \begin{subfigure}[b]{0.3\textwidth}
           \includegraphics[width=\textwidth]{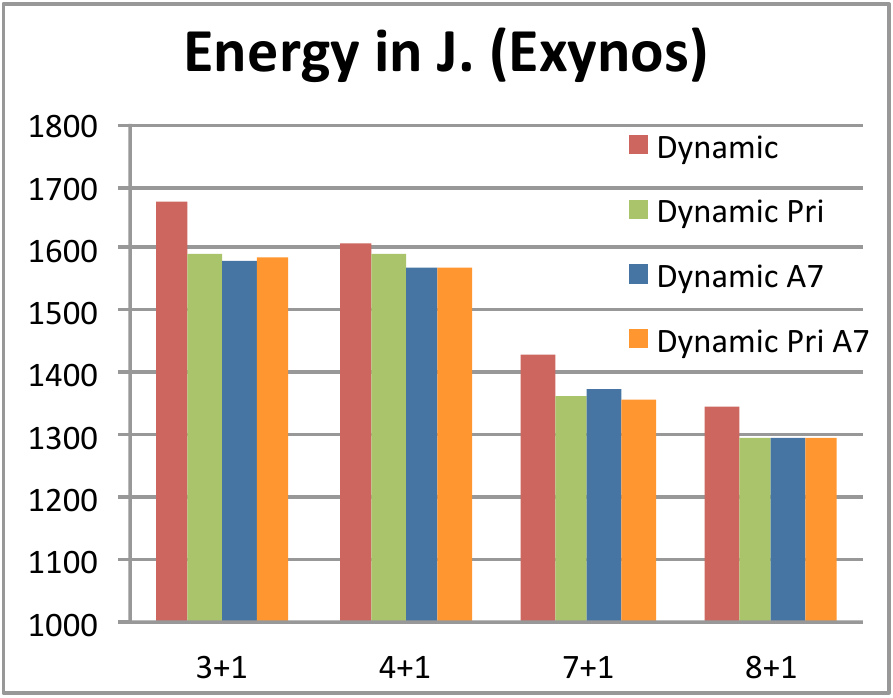}%
         \end{subfigure}
  ~~  \begin{subfigure}[b]{0.3\textwidth}
           \includegraphics[width=\textwidth]{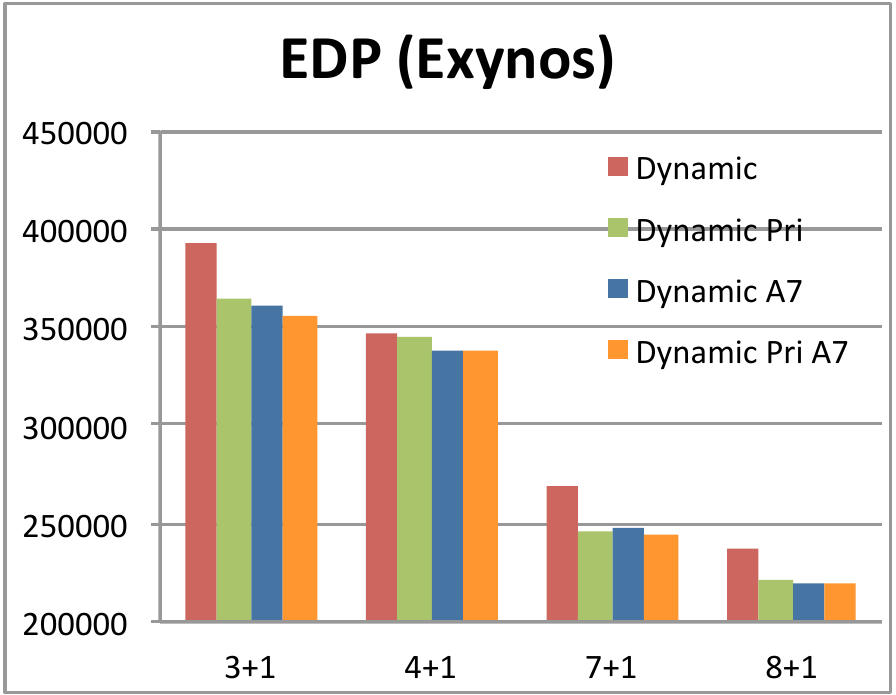}%
         \end{subfigure}
   \caption{Time (ms), energy (J) and energy-delay-product (EDP) for
     Barnes Hut on Exynos using A15 and A7 cores. The lower, the better.}\label{fig:BHbiglittle}
\end{figure*}

Finally, notice that incrementing the priority of the host thread does not
result in further reductions on time or energy on the Exynos
platform. However, in the next section we explore a posible strategy
to further reduce time and energy consumption on this architecture.

\subsection{Exploiting big.LITTLE architecture}
\label{big:little}

One interesting feature of the Exynos 5422 is that it supports global
task scheduling (GTS). This enables using the four A15 cores and the
four A7 ones at the same time. The Linux scheduler automatically uses
the more powerful A15 cores for compute intensive threads,
whereas power saving A7 are reserved for system and background tasks.

Figure~\ref{fig:BHbiglittle} shows execution time (ms.), energy (J.)
and EDP for the Exynos platform with configurations 3+1, 4+1, 7+1, and
8+1.  The 7+1 and the 8+1 configurations use both the A15 and A7 cores
for the computing threads, while the 3+1 and the 4+1 only use the A15
cores for the computing threads. In all the configurations, the
mapping of the host thread to an A15 or an A7 core depends on the
pinning strategy described below.  Notice that the GPU also computes
chunks of iterations, so there is now a larger degree of
heterogeneity. Four alternatives are considered for each
configuration:

\begin{itemize}
\item Dynamic is the baseline approach where the chunk size assigned
  to the A15 and A7 cores depends on the relative throughput on each
  one. The host thread is pinned to one of the A15 cores. 

\item Dynamic Pri is based on the baseline, but it increases
  the priority of the host thread. 

\item  Dynamic A7 is based on the baseline, but pins the host thread
  to one of the A7 cores.

\item Dynamic Pri A7 is based on the baseline, but it pins the 
host thread to one of the A7 cores and increases its priority.
\end{itemize}

The energy sampling thread (see Section~\ref{sec:meter}) is always
pinned to one of the A7 cores, so it only produces an overhead of
around 0.5\% of the total execution time when running Barnes Hut under
8+1 scenarios (less than 0.02\% otherwise). Regarding the breakdown of
energy consumption, A15 cores use between 72\% and 76\% of the total
energy consumed, the GPU between 15\% and 20\%, and the memory
subsystem around 4.5\%. Interestingly, A7 cores only take around 2.5\%
of the total energy when they are idle (3+1 and 4+1 scenarios) and
around  7.3\% when they are fully utilized (7+1 and 8+1
scenarios). For our benchmark, the four A7 cores consume one order of
magnitude less energy than the four A15 cores.

Results in Figure~\ref{fig:BHbiglittle} show that by increasing the
number of threads to use the A7 cores (compare 3+1 or 4+1 with 7+1 or
8+1), execution time and energy reduce significantly. In particular,
going from Dynamic 4+1 to Dynamic 8+1, we reduce time, energy and EDP
by 22\%, 19\% and 46\% respectively.  Increasing the priority of the host thread or pinning
it to a A7 have small impact.  For instance, for the 3+1 and the 4+1
configurations, increasing the priority of the host thread or pinning
it to the A7 core have almost no impact.  On the 4+1 scenario, where
Dynamic A7 is using 5 cores (4 A15 and 1 A7 for the host thread) we
appreciate a marginal energy saving of 3\% and similar running times.
Increasing the priority of the host thread or pinning it to the A7
core has a higher impact on the 7+1 and 8+1 configurations, although
their impact is relatively small.  We notice that for 7+1 and 8+1
configurations, Dynamic Pri, Dynamic A7, and Dynamic Pri A7, obtain
very similar results in terms of time and energy.  Again, for 7+1 and
8+1 scenarios and with respect to Dynamic, Dynamic Pri and Dynamic A7
can, on the average, reduce the time, energy and EDP by 4.3\%, 3.6\%
and 7.8\%, respectively. All in all, w.r.t. the Dynamic 4+1, 
Dynamic Pri 8+1 reduces EDP by 57\%.

Overall, while we expected that pinning the host thread to the A7
would have a higher impact on time or energy, our experimental results
show little impact on either of them.

\section{Related Work}\label{sec:related}

The closest work to ours is that of Zhu et al. \cite{zhuLCPC14},
which address the problem of performance degradation when several
independent OpenCL programs run at the same time (co-run) on the CPU
and on the GPU of an Ivy Bridge using the Windows OS. The programs
running on the CPU use all cores, so they are in a situation similar
to our 4+1 configurations (oversubscription).  To avoid degradation of
the GPU kernel they also propose increasing the priority of the thread
that launches the GPU kernel. Our study differs from theirs because we
do not run two different programs, instead we partition the iteration
space of a single program to exploit both, the CPU and the GPU.  Our
study also shows that increasing the priority of the host thread is
not necessary when there is no oversubscription (i.e. 3+1) or when the
underlying OS is Linux. We also assess the use of a big.LITTLE
architecture. 

Other works as \cite{Lustig, Grasso} also address the overhead
problems while offloading computation to GPUs. The work of Lustig and
Martonosi \cite{Lustig} presents a GPU hardware extension coupled with
a software API that aims at reducing two sources of overhead: data
transfers and kernel launching. They use a Full/Empty Bits technique
to improve data staging and synchronization in CPU-GPU
communication. This technique allows subsets of data results being
transferred to the CPU proactively, rather than waiting for the entire
kernel to finish. Grasso et al. \cite{Grasso} propose several host
code optimizations (Use of Zero-copy Buffer, Global Work Size equal to
multiples of \#EUs) in order to reduce GPU's computation overheads on
embedded GPUs. They present a comparison in terms of performance and
energy consumption between an OpenCL legacy version and an OpenCL
optimized one. Our work focus on reducing the sources of overhead as
well, but in contrast, we focus on CPU-GPU collaborative computation
instead of only targeting the integrated GPU.

Several previous works study the problem of automatically scheduling
on heterogeneous platforms with a multicore and an integrated or
discrete
GPU~\cite{Qilin,Ravi:hipc11,StarPU10,XKaapi2012,OmpSs12,Bel13,concord}.
Among those works, the only one that also uses chips with integrated
GPUs is Concord~\cite{concord}. However, Concord does not analyze the
overheads incurred by offloading a chunk of iterations to the GPU.

\section{Conclusions}\label{sec:conclusions}


In this work, we elaborate on the possibility of successfully
implementing a dynamic scheduling strategy that automatically
distributes the iteration space of a parallel loop among the cores and
the GPU of an heterogeneous chip. To that goal it is key to guarantee
optimal resource occupation and load balance while reducing the impact
of the overhead.  Our proposal is evaluated for the Barnes Hut
benchmark on mid/low power heterogenous architectures like Ivy Bridge,
Haswell, and Exynos 5, where the first two run under Windows OS and
the latter under Linux. We have studied the sources of overhead on
these systems, finding that, under Windows, the overhead due to
re-scheduling the host thread is prohibitive in oversubscribed
scenarios (more threads than cores). We solve this issue by increasing
the priority of the host thread. Our experimental results show that
this reduces Energy-Delay Product (EDP) by 18\% and 84\% on Intel Ivy
Bridge and Haswell architecture, respectively.  On the Exynos
platform, the Linux scheduler successfully deals with oversubscription
when only using the A15 cores.  Therefore, for this platform we
explore the benefits, in terms of time and energy, that can be
achieved when pinning the host thread to a low power core or by the
combined usage of the four A15 cores and the four A7 cores included in
the Exynos 5.  Our experimental results show that using the A7 cores
reduces EDP by 46\%. Increasing the priority of the host thread or
pinning the thread to an A7 reduces EDP by and additional 7.8\% in the
Exynos big.LITTLE architecture.


\bibliographystyle{abbrv}
\bibliography{parallelfor}  
\end{document}